\documentclass[aps,showpacs,twocolumn]{revtex4}
\usepackage{amsmath}
\usepackage{epsfig}

\begin{document}

\title{Configuration mixing and effective baryon-baryon interactions}

\author{Xinmei Zhu$^1$, Hongxia Huang$^1$, Jialun Ping$^1$\footnote{Corresponding
 author: jlping@njnu.edu.cn}, Fan Wang$^2$}

\affiliation{$^1$Department of Physics, Nanjing Normal University,
Nanjing 210097, P.R. China}

\affiliation{$^2$Department of Physics, Nanjing University,
Nanjing 210093, P.R. China}

\begin{abstract}
The effective baryon-baryon interactions is studied in the refined
quark delocalization color screening model (QDCSM), in which the different
quark clusterings are fully taken into account, instead of controlling by
a variational delocalization parameter $\epsilon(s)$ between two
3-quark clusters. The symmetry bases are employed to do the calculation,
all possible configurations for two quark clusters are considered.
The results obtained are very similar to that of QDCSM. It is inferred that
the delocalization parameter $\epsilon(s)$ used in QDCSM is an economic
and effective way to describe the mixing of quarks between baryons.
\end{abstract}

\pacs{12.39.Jh, 13.75.Cs, 14.40.Rt}

\maketitle

\section{Introduction}

Since the advent of quantum chromodynamics (QCD), the fundamental
theory of strong interaction, it has been expected to calculate the
properties of baryons and baryon-baryon (BB) interactions from QCD directly.
However, the complexity of the non-perturbative character of QCD in the
low energy region has hindered this attempt. Although the lattice QCD~\cite{LQCD},
Dyson-Schwinger equation approach~\cite{DSE}, QCD sum rule~\cite{sumrule},
chiral perturbation theory~\cite{chiral} and other non-perturbative methods
have made impressive progresses, the quark model approach is still the
important method in hadron physics, especially for BB interaction and
multiquark system. The quark model describes the experimental data
of hadron properties and hadron-hadron interactions very well.

The naive quark model (Glashow-Isgur model~\cite{GI}) successfully described
the properties of baryons. However to extend it to BB interaction,
refinements are necessary. One approach is to add the Goldstone-boson-exchange
interaction to the Hamiltonian, it is developed to the chiral quark model~\cite{chiralQM}.
After fine-tuning the model parameters, the model give a satisfactory
description of BB interactions, where the $\sigma$-meson and other scalar mesons
play an indispensable role. However, the $\sigma$-meson, as an $S$-wave resonance of $\pi\pi$
\cite{BESsigma}, cannot provide the expected intermediate-range attraction for nucleon-nucleon
interaction~\cite{sigma}. In addition, how to realize the chiral partner
in flavor $SU_3$ case is still a problem. The outstanding feature of the similarity
between the nuclear force and molecular force can not have an explanation in
this meson exchange model approach. The another approach is the
quark delocalization color screening model (QDCSM)~\cite{QDCSM}. Two
ingredients, quark delocalization and color screening, are introduced
to enlarge the model space and modify the interactions of quark-pair
in different quark clusters. This model is based on the same idea as
Heitler and London's approach of hydrogen molecular structure.
It explains the similarity between molecular force and nuclear
force naturally. It also gives a good description of the
BB interaction. Its predictions on $d^*$ dibaryon~\cite{QDCSM1} had been confirmed by
Celsius-WASA experiments~\cite{dstar}, the prediction of $N\Omega$ dibaryon~\cite{NOmega} is also
supported by lattice QCD calculation~\cite{LQCD2}. In QDCSM, the color screening is used
to lower the color confinement potential between two baryons and to facilitate the
quark delocalization. The further studies show that the color screening phenomenology
is an effective description of the hidden-color channels coupling~\cite{color}.
The quark delocalization is realized by introducing a delocalization parameter $\epsilon(s)$,
which is determined by the dynamics of the system,
\begin{eqnarray}
& & \psi_l(\mathbf{r}) =
\left({\phi}_L(\mathbf{r}) +\epsilon {\phi}_R(\mathbf{r}) \right) /N(\epsilon),
\nonumber \\
& & \psi_r(\mathbf{r}) =
\left({\phi}_R(\mathbf{r}) +\epsilon {\phi}_L(\mathbf{r}) \right) /N(\epsilon),
 \nonumber \\
& & N(\epsilon)=\sqrt{1+2\epsilon v+\epsilon^{2}},~~~~
v=\langle\phi_L|\phi_R\rangle, \nonumber
\end{eqnarray}
Generally the un-delocalized single particle wavefunctions take the Gaussian form,
\begin{eqnarray}
|L\rangle \equiv {\phi}_L(\mathbf{r}) & = &
\left(\pi b^2\right)^{-\frac{3}{4}}
e^{-\frac{1}{2b^2}{(\mathbf{r}+\mathbf{S}/2)}^2}, \nonumber \\
|R\rangle \equiv {\phi}_R(\mathbf{r}) & = &
\left(\pi b^2\right)^{-\frac{3}{4}}
e^{-\frac{1}{2b^2}{(\mathbf{r}-\mathbf{S}/2)}^2}
\end{eqnarray}

In fact, the full delocalization can be realized
by configuration mixing, i.e., taking all the configurations $|L^6\rangle,
|L^5R\rangle, |L^4R^2\rangle, |L^3R^3\rangle, |L^2R^4\rangle, |LR^5\rangle$
and $|R^6\rangle$ into consideration for BB interaction, under the two
cluster approximation. The mixing of the above seven configurations is
determined by the dynamics of the system directly
and no longer by one delocalization parameter $\epsilon$. In this way, the model space
is larger than that of QDCSM. Fl. Stancu and L. Wilets had pointed that all the
configuration are important in their studies of nucleon-nucleon interaction~\cite{wilets}.
So it is interesting to study the BB interactions
with the configuration mixing method and to check the validity and efficiency of
delocalization method in QDCSM. For simplicity, the first version of QDCSM, where
the interaction between quark pair consists of color confinement and one-gluon-exchange
only, is employed in the present study.

This paper is organized as follows. In section II, the model
Hamiltonian, the symmetry bases and the calculation method are described.
The results are given in section IV and a summary is given in the last section.

\section{Models and bases}

The model Hamiltonian is the same as that of QDCSM, which is described in detail in
Ref.\cite{QDCSM}. Here we only write down the Hamiltonian for 6-quark system,
\begin{eqnarray}
H(6) & = & \sum_{i=1}^6 (m_i+\frac{\mathbf{p}_i^2}{2m_i})
+\sum_{i<j=1}^{6}V_{ij} -T_c(6),        \label{ham} \\
V_{ij} & = & V_{ij}^c + V_{ij}^G, \nonumber \\
V_{ij}^c & = & -a_c \boldsymbol{\lambda}_i\cdot \boldsymbol{\lambda}_j
 \left\{ \begin{array}{ll}  r_{ij}^2 & ~~\mbox{if }i,j\mbox{ occur in the} \\
 &  ~~\mbox{same baryon orbit}, \\
\frac{1-e^{-\mu r_{ij}^2}}{\mu} & ~~\mbox{if }i,j\mbox{ occur in }\\  &
 ~~\mbox{different baryon orbits}.
\end{array}
\right.
  \nonumber \\
V_{ij}^G & = & \alpha_s \frac{ \boldsymbol{\lambda}_i \cdot \boldsymbol{\lambda}_j}{4}
 \left[ \frac{1}{r_{ij}}-\frac{\pi}{2}\delta (\mathbf{r}_{ij}) \left( \frac{1}{m_i^2}+\frac{1}{m_j^2}
 +\frac{4\boldsymbol{\sigma}_i \cdot \boldsymbol{\sigma}_j}{3m_im_j}  \right)
 \right]. \nonumber
\end{eqnarray}

To do the calculation, the symmetry bases are more convenient. The symmetry bases
are the group-chain classification bases, which are defined as follows~\cite{prc511}
\begin{equation}
\Phi^{\alpha}_{Kmn} (q^6) = \left | \begin{array}{c}
[\nu]L^mR^n \\ ~[\sigma]W[\mu]\beta[f]YIJM_IM_J \end{array} \right>.
\end{equation}
where $\alpha=(YIJ)$ are the strong interaction conserved quantum numbers:
strangeness, isospin and spin. $K$ denotes the intermediate quantum numbers,
$[\nu],[\mu],[f]$. $[\nu],[\mu],[\sigma],[f]$ represent the symmetry of orbital,
spin-flavor $SU(6)$, $SU(3)$ color and flavor. $[\sigma]=[222]$ is fixed due to
the color singlet requirement.
For some interesting sets of quantum numbers $\alpha$, the
allowed intermediate quantum numbers $K$ are listed in Table \ref{K}.
\begin{table}
\caption{The allowed $K$ for interesting sets of $\alpha$.
The indices of the symmetries $[\nu],[\mu], [f]$ are: 1-[6]; 2-[51]; 3-[42];
4-[33]; 5-[411]; 6-[321]; 7-[222]; 8-[3111]; 9-[2211]. \label{K} }
\begin{tabular}{ccccccccccc} \hline
$\alpha$ & \multicolumn{7}{c}{$K$} \\ \hline
201 & 144 & 264 & 324 & 344 & 394 & 364 & 354 & & & \\
\hline
210 & 143 & 263 & 323 & 343 & 393 & 363 & 353 & & & \\
\hline
203 & 144 & 344 & & & & & & & & \\
\hline
$-400$ & 141 & 341 & & & & & & & & \\
\hline
~~$-1\frac{1}{2}2$~~ & 143 & 146 & 233 & 232 & 235 & 236 & 263 & 266 & 265 & 323 \\
                 & 322 & 325 & 353 & 356 & 352 & 355 & 343 & 346 & 363 & 366 \\
                 & 365 & 396 & 412 & 433 & 436 & 432 & 435 & 486 & 485 & 476 \\ \hline
\end{tabular}
\end{table}
Table \ref{nu} presents the allowed spatial symmetries $[\nu]$ for given $m,n$.
The number of symmetry bases for given $\alpha$ can be obtained from the
combination of the allowed $K$ and $m,n$, which is given in Table \ref{nbases}.
\begin{table}
\caption{the allowed spatial symmetry for given $m,n$. \label{nu}}
\begin{tabular}{cccccccc} \hline
 {$[\nu]$} &$L^6$ &$L^5R$ &$L^4R^2$ &$L^3R^3$ &$L^2R^4$ &$LR^5$ &$R^6$ \\
\hline
[6] & 1 & 1 & 1 & 1 & 1 & 1 & 1 \\
\hline
[51] & 0 & 1 & 1 & 1 & 1 & 1 & 0 \\
\hline
[42] & 0 & 0 & 1 & 1 & 1 & 0 & 0 \\
\hline
[33] & 0 & 0 & 0 & 1 & 0 & 0 & 0 \\
\hline
\end{tabular}

\caption{The number of symmetry bases for interesting physical channels. \label{nbases}}
\begin{tabular}{cccccc}
\hline
 {YIJ} &201 &210 &203 &-400 &-1$\frac{1}{2}$2 \\
\hline
 {coupling channels} & 27 & 27 & 10 & 10 & 96  \\
\hline
\end{tabular}
\end{table}

For illustration, two bases of the ten bases for channel $\alpha=(YIJ)=203$ are shown below,
\begin{eqnarray}
\Phi^{(203)}_{144~51} (q^6) & = & \left | \begin{array}{c}
[6]L^5R^1 \\ ~[222]1[33]1[33]20303 \end{array} \right>, \nonumber \\
\Phi^{(203)}_{344~24} (q^6) & = & \left | \begin{array}{c}
[42]L^2R^4 \\ ~[222]1[33]1[33]20303 \end{array} \right>. \nonumber
\end{eqnarray}
where $M_I,M_J$ take their maximum values, the eigenenergies are independent of
$M_I,M_J$.

To study the BB interaction, the Schr\"{o}dinger equation for
6-quark system has to be solved,
\begin{equation}
H(6)\Psi^{\alpha}(q^6)=E^{\alpha}\Psi^{\alpha}(q^6), \label{sch}
\end{equation}
where the eigen wavefunction $\Psi^{\alpha}$ is the linear combination of
$\Phi^{\alpha}_{Kmn}(q^6)$ under the cluster approximation,
\begin{equation}
\Psi^{\alpha}(q^6)=\sum_{Kmn} c^{\alpha}_{Kmn}\Phi^{\alpha}_{Kmn}(q^6). \label{combine}
\end{equation}
By using Eq.(\ref{combine}), Eq.(\ref{sch}) becomes
\begin{equation}
\sum_{k'} \left[ \langle \Phi^{\alpha}_{k'}|H(6)|\Phi^{\alpha}_{k}\rangle
-E^{\alpha} \langle \Phi^{\alpha}_{k'}|\Phi^{\alpha}_{k}\rangle \right]
c^{\alpha}_{k}(q^6)= 0,  \label{eigen}
\end{equation}
where $k$ stands for $Kmn$. The eigen energy of the system can be obtained
by solving the generalized eigen equation. the calculation of
the 6-quark Hamiltonian matrix elements on the symmetry basis is performed
by the well known fractional parentage expansion technique~\cite{prc511}.
\begin{eqnarray}
\langle \Phi^{\alpha}_{k}|H| \Phi^{\alpha}_{k'} \rangle
 & = & \sum \left(^6_2 \right)
\langle \Phi^{\alpha}_{k} | \alpha_1 k_1, \alpha_2 k_2 \rangle
\langle \alpha^{\prime}_1 k^{\prime}_1, \alpha^{\prime}_2
k^{\prime}_2 | \Phi^{\alpha}_{k'} \rangle  \nonumber \\
& &
\langle \alpha_1 k_1 | \alpha^{\prime}_1 k^{\prime}_1 \rangle
\langle \alpha_2 k_2 | H_{56} | \alpha^{\prime}_2 k^{\prime}_2 \rangle .
\label{cfp}
\end{eqnarray}
where $\langle \alpha_1 k_1 | \alpha^{\prime}_1 k^{\prime}_1 \rangle$
is the four quark overlap. $\langle \alpha_2 k_2 | H_{56} |
\alpha^{\prime}_2 k^{\prime}_2 \rangle$ is the two body matrix element
and $H_{56}$ represents the two-body operator for the last pair.
$\langle \Phi^{\alpha}_{k} | \alpha_1 k_1, \alpha_2 k_2 \rangle$
and $\langle \alpha^{\prime}_1 k^{\prime}_1, \alpha^{\prime}_2 k^{\prime}_2
| \Phi^{\alpha}_{k^{\prime}}\rangle$ are the total coefficients of
fractional parentage. All the needed coefficients can be obtained from
the Chen's book~\cite{book}.

\begin{figure}
\epsfxsize=3.5in \epsfbox{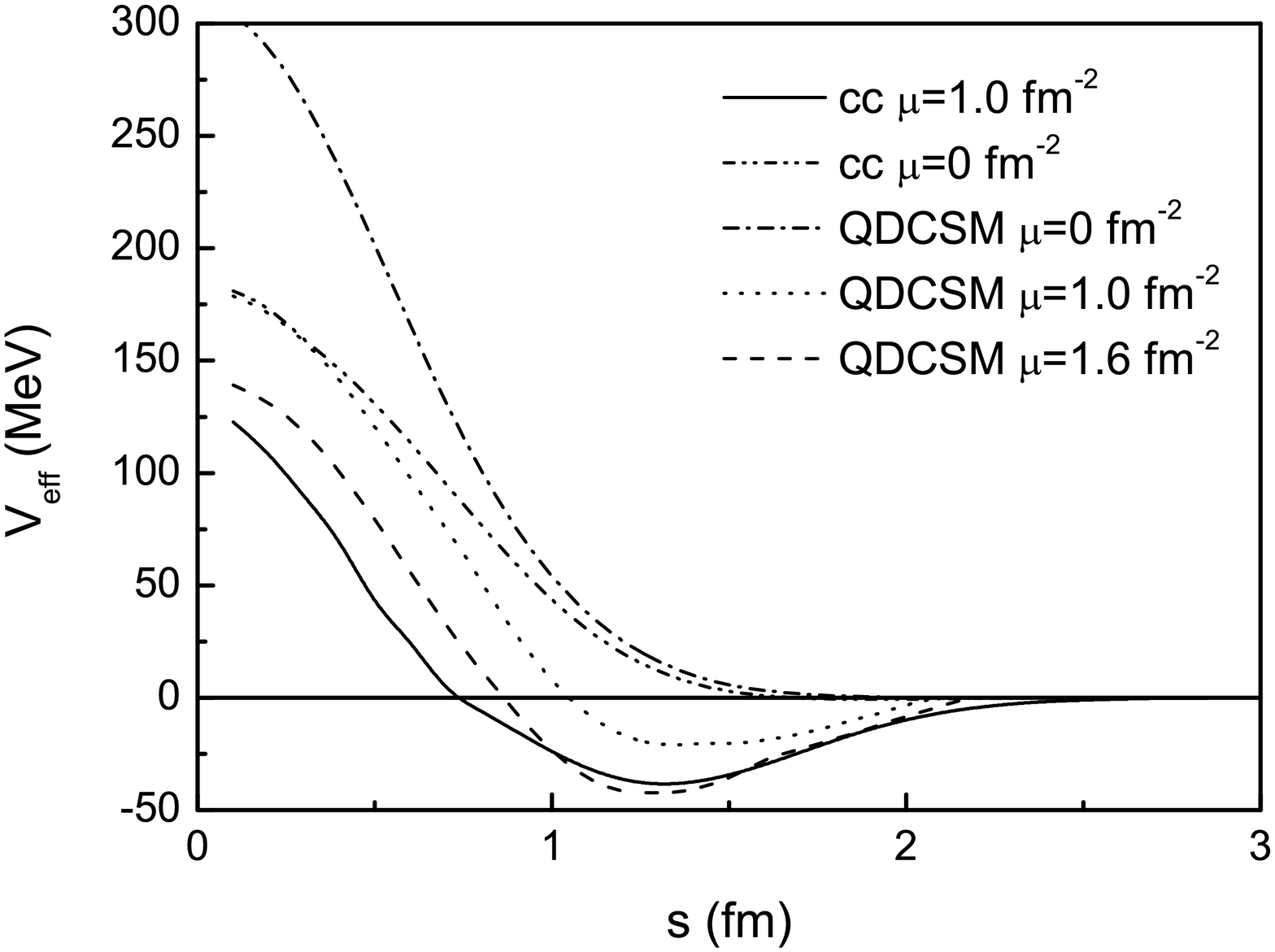}
\caption{The effective potentials for $NN$ with $YIJ=201$. 
'cc' stands for configuration mixing calculation.}
\end{figure}

\section{Results}
The effective potentials between two baryons in the $u,d,s$ 3-flavor
world are calculated in the framework of the refined QDCSM.
The effective potential is defined as
\begin{equation}
V_e=E_6(s)-E_6(S=\infty)
\end{equation}
where $E_6(s)$ is the eigen-energy of 6-quark system with separation $s$.
To save space, only the results of several interesting states are given below.
The needed model parameters are all taken from QDCSM, which is listed in
Table \ref{parameter}.
\begin{table}
\caption{The parameters used in the calculations.\label{parameter}}
\begin{tabular}{cccccc} \hline
 $m$(MeV) & $m_s$(MeV) & $b$(fm)
  & $\alpha_s$ & $a_c$(MeV$\cdot$fm$^{-2}$) & $\mu$(fm$^{-2}$)  \\ \hline
 313   & 633.76  & 0.603   & 1.54  & 25.13 &  0 or 1.0  \\ \hline
\end{tabular}
\end{table}

Before presenting the results, we discuss the problem of numerical calculation
first. When the separation between two clusters is very small, the difference
between $L$ and $R$ goes to vanishing, the set of bases of the system
will be over complete. To remove the spurious bases, the eigen method is
used. First the overlap matrix of the system is diagonized, the eigenvectors
corresponding to zero eigenvalues are dropped. Then the Hamiltonian matrix
is reproduced on the remained eigenvectors of overlap matrix with non-zero eigenvalues.
At last the new Hamiltonian matrix is diagonized to get the eigen-energy of the system.

Another problem needs to mention is associated with two configurations $L^6$ and $R^6$.
Physically, these two configurations are the same, and the energies of these two
configurations do not change with the separation between two clusters.
In fact, these two configurations almost do not contribute to the effective potentials.

Figs. 1-5 give the effective potentials between two baryons with
quantum numbers $YIJ=201$, $210$, $203$, $-1\frac{1}{2}2,-400$.
For comparison, the results for $\mu=0$ fm$^{-2}$ (without color screening)
and the results of QDCSM are also shown in these figures.
Clearly, the effective potentials go to zero when the separation becomes
large (e.g., $s=3$ fm),
due to characteristic of confinement and one-gluon-exchange potentials
and the Gaussians used as the wavefunctions of single-particle.
All figures also show the differences between the present approach and QCDSM
gradually disappear when the separation becomes large, because the color
confinement pushes the energies of the hidden color configurations,
$[6]R^5L^1$, $[6]R^4L^2$, $[6]R^2L^4$, $[6]R^1L^5$, 
$[42]R^4L^2$, $[42]R^2L^4$, higher.

Fig. 1 shows the effective potentials between two nucleons with quantum
numbers $YIJ=201$. If there is no color screening effect, i.e.,
$\mu=0$ fm$^{-2}$, then the intermediate range attraction is missing both
in present work and in QDCSM, and QDCSM have a larger repulsive core. Taking into
account of the color screening effect, the intermediate range attraction
is obtained in both approaches. The attraction in QDCSM is about half of
that in the present approach. This is reasonable, because the variational
space in the present approach is larger than that in QDCSM.
Even though all the configurations, $|L^6\rangle,
|L^5R\rangle, |L^4R^2\rangle, |L^3R^3\rangle, |L^2R^4\rangle, |LR^5\rangle$
and $|R^6\rangle$ are included in QDCSM, the percentages
of different configurations are controlled by one
variational parameter $\epsilon$, which is determined by system dynamics.
Whereas in the refined QDCSM the dynamics of system fix the configuration
mixing with 7 coefficients (see Table V). By increase the color screening
parameter, e.g., $\mu=1.6$ fm$^{-2}$,
QDCSM can produce almost the same effective potentials as the present
approach.

\begin{figure}
\epsfxsize=3.5in \epsfbox{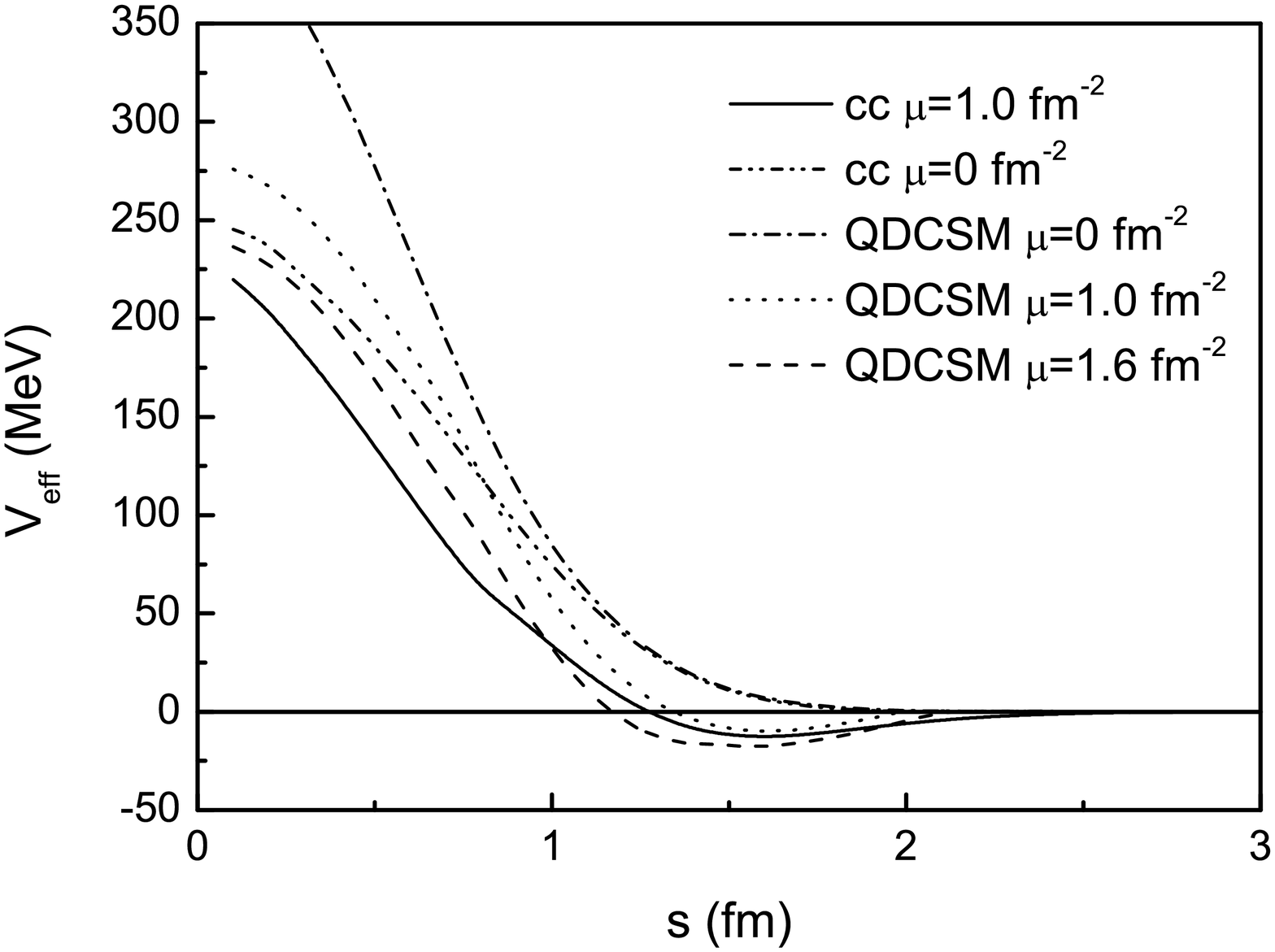}
\caption{The same as Fig.~1 for $NN$ with $YIJ=210$.}
\end{figure}

\begin{figure}[b]
\epsfxsize=3.5in \epsfbox{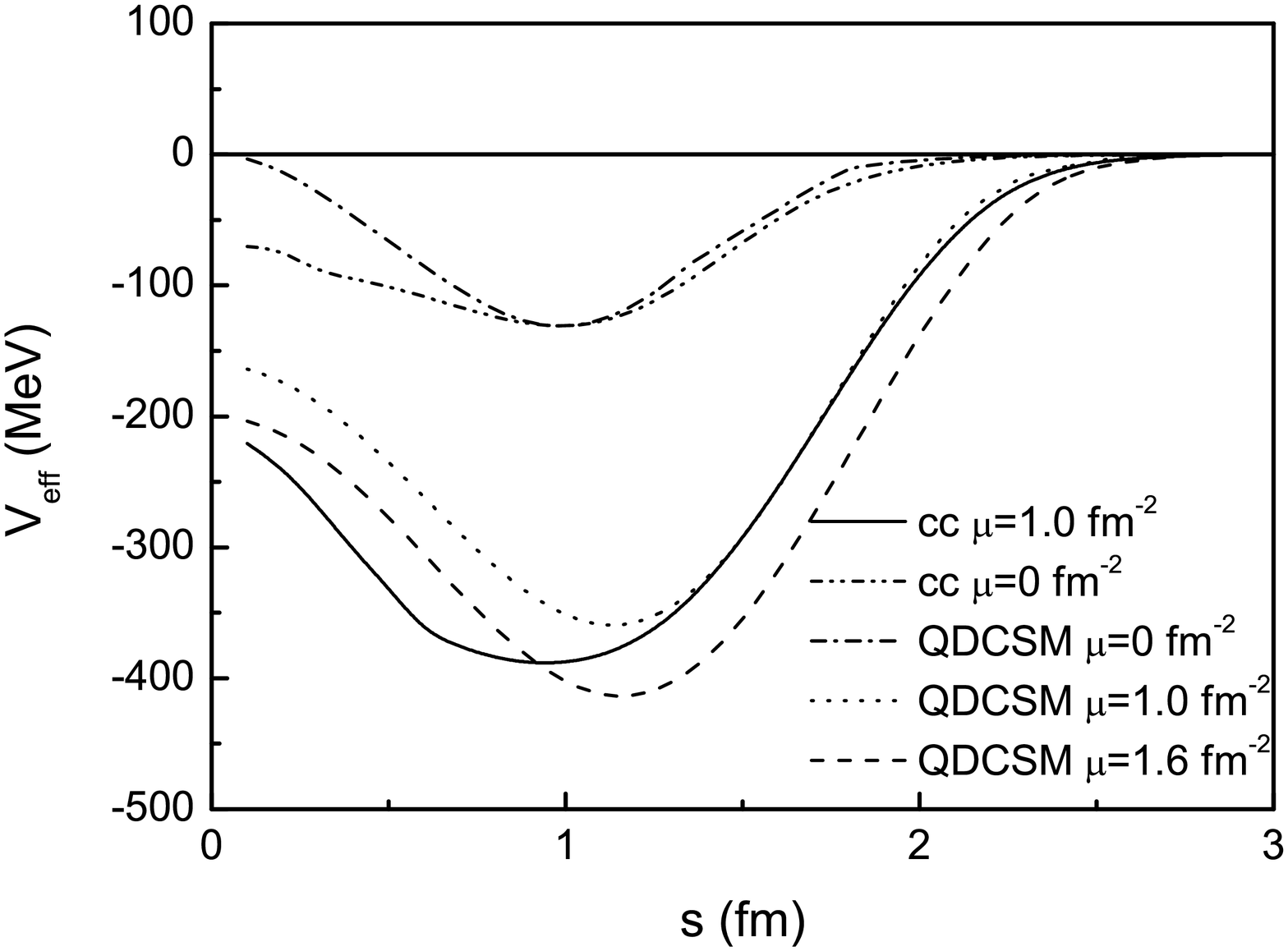}
\caption{The same as Fig.~1 for $NN$ with $YIJ=203$.}

\epsfxsize=3.5in \epsfbox{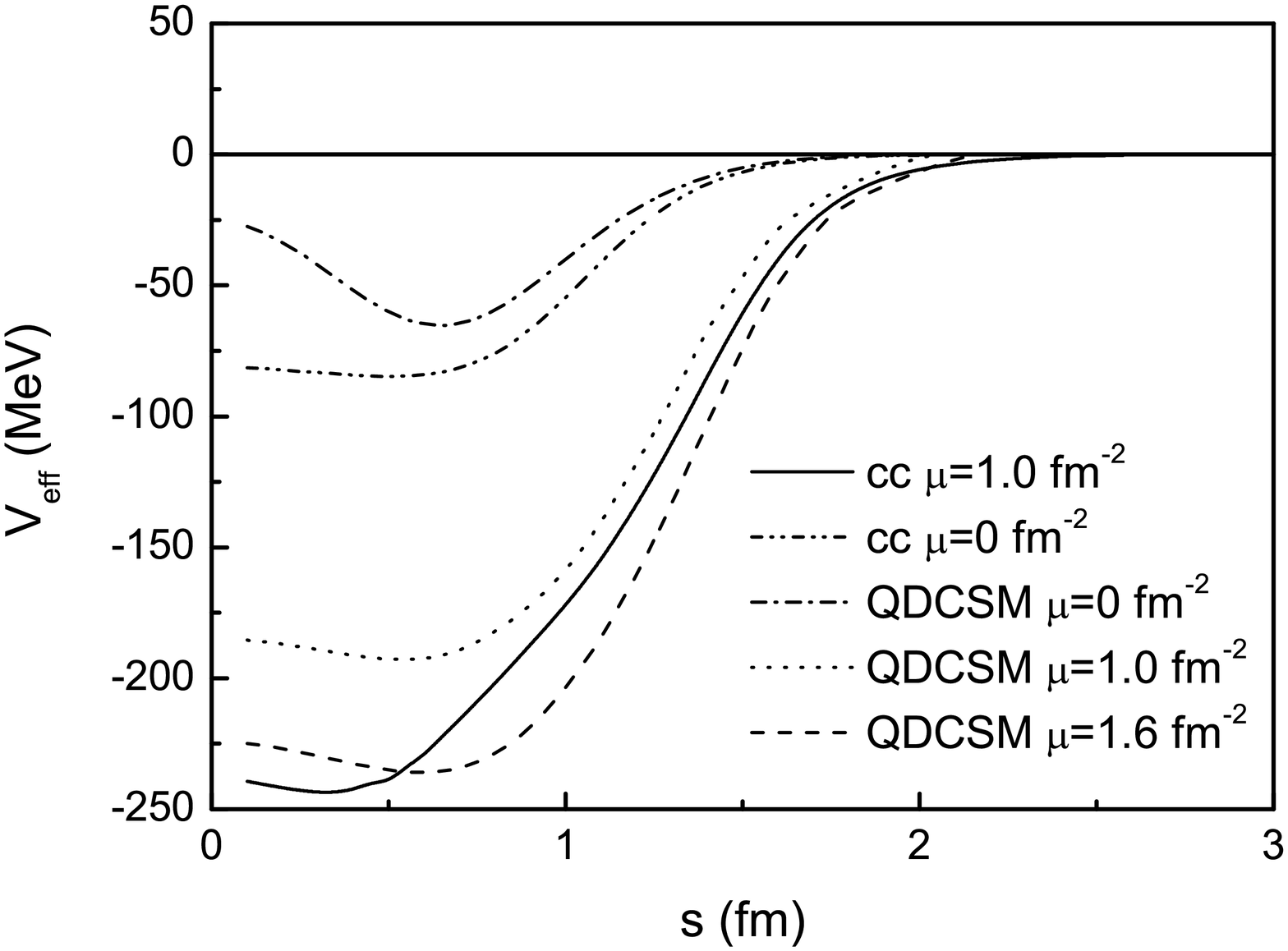}
\caption{The same as Fig.~1 for $N\Omega$ with $YIJ=-1\frac{1}{2}2$.}

\end{figure}

\begin{table*}[ht]
\caption{The relative probabilities of different configurations in the states for
$YIJ=203$ with $\mu=1.0$ fm$^{-2}$. \label{probability}}
\begin{tabular}{c|cc|cc|cc|cc|cc} \hline
 & \multicolumn{2}{c|}{$s=4$ fm} &  \multicolumn{2}{c|}{$s=2.6$ fm} &  \multicolumn{2}{c|}{$s=2$ fm}
 & \multicolumn{2}{c|}{$s=1.1$ fm} &  \multicolumn{2}{c}{$s=0.5$ fm}  \\ \hline
 & This work  & QDCSM  &  This work  & QDCSM  &  This work  & QDCSM  &  This work  & QDCSM &
   This work  & QDCSM  \\ \hline
 $[6]R^6$    & 0.00 & 0.00 & 0.33 & 0.00 & 0.27 & 0.01 & 0.00 & 0.02 & 0.00 & 0.00 \\
 $[6]R^5L^1$ & 0.00 & 0.00 & 0.01 & 0.00 & 0.47 & 0.14 & 0.00 & 0.19 & 0.05 & 0.11 \\
 $[6]R^4L^2$ & 0.00 & 0.00 & 0.04 & 0.11 & 0.74 & 0.61 & 0.25 & 0.67 & 0.08 & 0.59 \\
 $[6]R^3L^3$ & 1.00 & 1.00 & 1.00 & 1.00 & 1.00 & 1.00 & 1.00 & 1.00 & 1.00 & 1.00 \\
 $[6]R^2L^4$ & 0.00 & 0.00 & 0.04 & 0.11 & 0.74 & 0.61 & 0.25 & 0.67 & 0.08 & 0.59 \\
 $[6]R^1L^5$ & 0.00 & 0.00 & 0.01 & 0.00 & 0.47 & 0.14 & 0.00 & 0.19 & 0.05 & 0.11 \\
 $[6]L^6$    & 0.00 & 0.00 & 0.33 & 0.00 & 0.27 & 0.01 & 0.00 & 0.02 & 0.00 & 0.00 \\
 $[42]R^4L^2$ & 0.00 & 0.00 & 0.02 & 0.07 & 0.12 & 0.09 & 0.01 & 0.00 & 0.01 & 0.00 \\
 $[42]R^3L^3$ & 4.00 & 4.00 & 3.79 & 3.29 & 0.61 & 0.36 & 0.01 & 0.00 & 0.00 & 0.00 \\
 $[42]R^2L^4$ & 0.00 & 0.00 & 0.02 & 0.08 & 0.12 & 0.09 & 0.01 & 0.00 & 0.01 & 0.00 \\
 \hline
\end{tabular}
\end{table*}

The effective potentials for $NN$ with $YIJ=210$ are shown in Fig. 2.
The results of comparison between two approaches are similar to that of
$NN$ with $YIJ=201$.

For $YIJ=203$ $\Delta$-$\Delta$ channel, attractions appear in all cases (see Fig. 3).
The color screening introduces much stronger attraction. With the same color
screening parameter, the present approach give a little stronger attraction
than QDCSM in the short-range part. Increasing the color screening parameter in QDCSM,
the curve will move downward, getting close to the results of the present
approach. Due to the strong attraction, the bound state with respect to two $\Delta$'s
can be formed. Taking into account the four-body decay channel, $NN\pi\pi$, a resonance
appears in this channel, which was observed in WASA-at-COSY experiments~\cite{dstar}.
To show the reason for the similarity between two approaches in detail, the relative
probabilities of different configurations in the given states, the ratios of probability of
a configuration to the configuration $[6]R^3L^3$, are tabulated in Table V. Due to the
symmetry between left and right Gaussians, the probabilities of the configuration
$[\nu]R^mL^n$ in the states are identical to that of $[\nu]R^nL^m$. From this table,
we can see that the same physical states are obtained in the two approaches when the
separation is large enough ($s=4$ fm). With the decreasing of the separation, the differences
between the two approaches appear gradually. However, the main configurations in the
states are kept the same, $[6]R^3L^3,~[42]R^3L^3$ and $[6]R^4L^2,~[6]R^2L^4$. Even as the
separation becomes very small, the superficial large difference between two approaches
will be reduced remarkably because of the large
overlap among the configurations $[6]R^mL^n$ with different $m,n$, due to the
fact that there is no much difference between the right and left gaussians.

\begin{figure}[ht]
\epsfxsize=3.5in \epsfbox{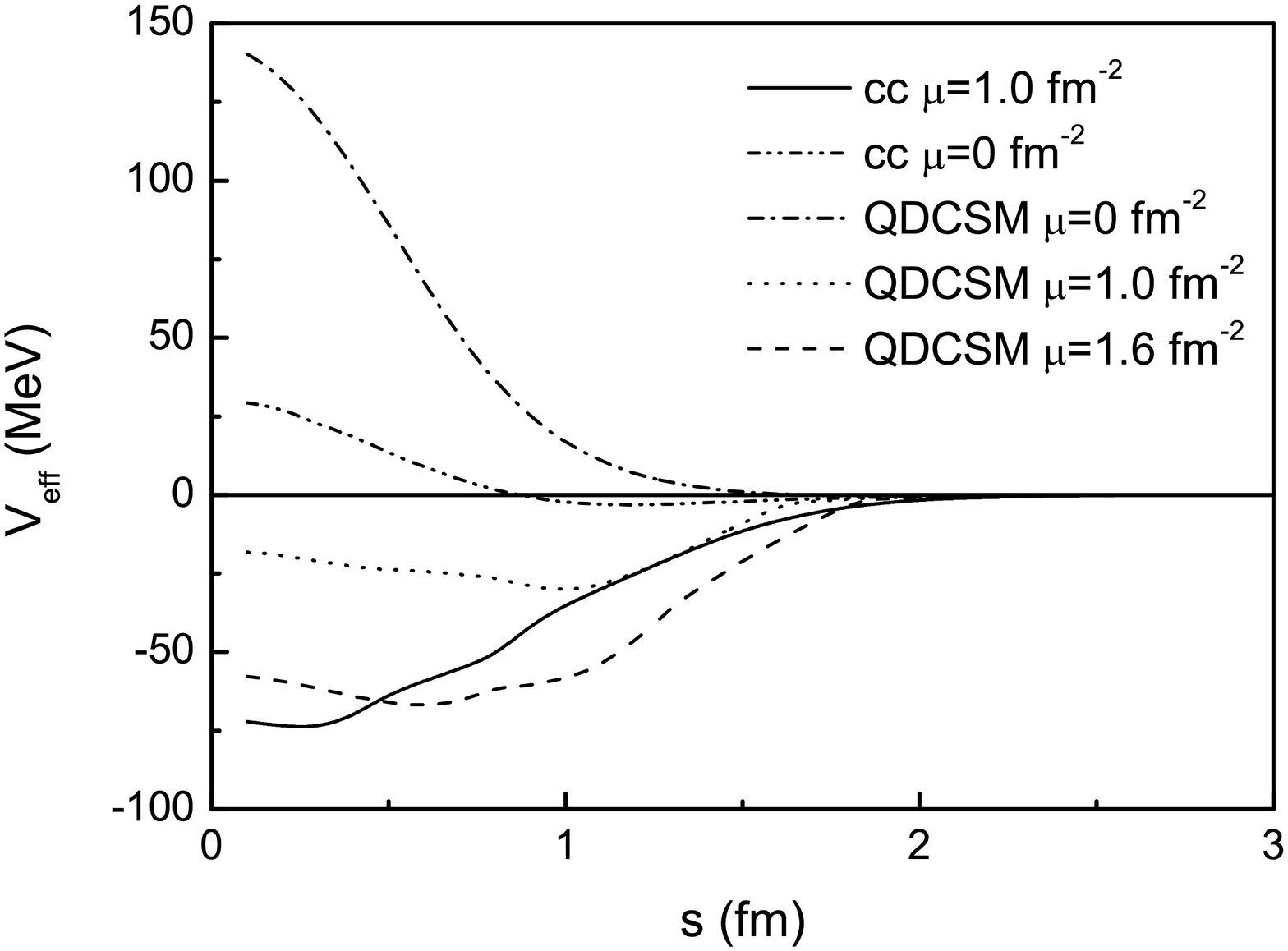}
\caption{The same as Fig.~1 for $\Omega\Omega$ with $YIJ=-400$.}
\end{figure}

The effective potentials of $N\Omega$ with $YIJ=-1\frac{1}{2}2$ are given
in Fig. 4. Again, attractions appear in all cases, the color screening increases
the attraction about 100 MeV at the short-range part. A resonance is expected
to be formed in this channel because of the deep attraction. The dynamical
calculation of quark model and the lattice QCD calculation supported this result.

Di-$\Omega$ as a possible dibaryon candidate was predicted in 1990~\cite{diOmega}
and also proposed by Li {\em et al.} in 2001~\cite{Li}. From Fig. 5, we can see that
the effective potential between two $\Omega$'s has a mild attraction in QDCSM with
color screening $\mu=1$ fm$^{-2}$. However, a rather strong attraction in the
short-range part is obtained in the present approach with the same $\mu$.
Taking into account the fact that the two $\Omega$'s can form a weak bound state
in QDCSM, it is expected that the $\Omega\Omega$ with $YIJ=400$ is a good
dibaryon candidate.

\section{Summary}
By enlarging the model space, i.e., taking into account of all possible configurations
under the constraint of two clusters, the effective potentials are obtained
for the BB systems. In most cases, the potentials obtained in the present approach
are similar to the ones obtained in QDCSM, except at the short-range part, where
lower potentials obtained in this approach. The results show the validity of the
delocalization in QDCSM.

From the results for NN channels, the configuration mixing themselves does not lead to the
intermediate-range attraction unless the color screening effect is introduced.
The possible dibaryon candidates based on the present results are similar to that of
the previous work~\cite{prc511}. Maybe $\Omega\Omega$ will have a little large
binding energy in the present approach. To obtain more reliable results for dibaryons,
the dynamical calculation is indispensable, which is in progress.

\acknowledgments{This work is supported partly by the National
Science Foundation of China under Contract Nos. 11175088,
11035006, 11265017}

\end{document}